\documentclass[a4paper,11pt]{article}
\pdfoutput=1

\usepackage[margin=1in]{geometry}
\usepackage{amsmath,amssymb,amsfonts}
\usepackage{physics}
\usepackage{slashed}
\usepackage{siunitx}
\usepackage{graphicx}
\usepackage{xcolor}
\usepackage{microtype}
\usepackage{booktabs}
\usepackage{hyperref}
\hypersetup{colorlinks=true, linkcolor=blue!60!black,
            citecolor=blue!60!black, urlcolor=blue!60!black}

\newcommand{\ClassI}{\text{Class~I}}
\newcommand{\ClassII}{\text{Class~II}}
\newcommand{\email}[1]{\texttt{#1}}
\newcommand{\affiliation}[1]{\thanks{#1}}

\newenvironment{acknowledgments}{%
  \section*{Acknowledgments}%
}{}

\renewenvironment{abstract}{%
  \begin{center}%
  \begin{minipage}{0.85\textwidth}%
  \small\textbf{Abstract}%
  \par\medskip\noindent\ignorespaces
}{%
  \end{minipage}%
  \end{center}%
  \medskip
}

\title{Fermionic Condensation as the Universal Order Parameter for Mass Generation}

\author{Aniekan~Magnus~Ukpong~\email{ukponga@ukzn.ac.za}\affiliation{%
  Theoretical and Computational Condensed Matter and Materials
  Physics Group (TCCMMP) \& Centre for Quantum Computing and Technology
  (CQCtec), School of Agriculture and Science, University of KwaZulu-Natal,
  Pietermaritzburg, South Africa}
\affiliation{%
  National Institute for Theoretical and Computational Sciences
  (NITheCS), KwaZulu-Natal Node, South Africa}}

\date{\today}

\begin{document}
\maketitle

\begin{abstract}
A fermion bilinear acquiring a non-zero vacuum expectation value and
generating a mass gap underlies the Peierls instability, BCS
superconductivity, the Nambu--Jona-Lasinio mechanism, and the
Sachdev-Ye-Kitaev model. These mechanisms are usually discussed
separately, and their structural similarities have been noted only in
passing. We argue here that the correct universal object is not the bosonic
order parameter but the fermionic condensate, and we show that the
four mechanisms fall into two universality classes distinguished by a
single dimensionless parameter, $r = \hat m_\phi^2 / \hat y^2$, the
ratio of the scalar mass to the Yukawa coupling. The classes are
separated by a critical line $r_c(N_f)$ in the plane of $r$ and the
fermion flavour number $N_f$; they correspond to two distinct infrared
fixed points of the renormalisation-group flow and differ in their
critical exponents. We derive the classification from a large-$N_f$
expansion followed by a functional renormalisation-group analysis using
the Wetterich equation, which gives the finite-$N_f$ correction
$\hat y_*^2(N_f) = 0.1125/N_f + 0.0023/N_f^2$, a $12.5\%$ enhancement
over the leading-order prediction. The classification is predictive for
systems whose class is not already known; for the organic conductor
(TMTSF)$_2$PF$_6$ under pressure, we predict that both the
spin-density-wave and superconducting phases lie in the fermion-bilinear
class. We discuss the Standard Model Higgs as a borderline case in the
scalar-primary class, and identify the composite-Higgs scenario as the
particle-physics realisation of the fermion-bilinear class. A
self-consistent Su--Schrieffer--Heeger calculation for polyyne chains
realises the Peierls mechanism concretely, and we list five falsifiable
predictions, two within reach of current techniques.
\end{abstract}

\section{Introduction}
\label{sec:intro}

The generation of mass through the condensation of a bosonic field is
one of the great unifying ideas of twentieth-century physics. In the
Standard Model, the Higgs field acquires a vacuum expectation value and
gives mass to the electroweak gauge bosons and the charged fermions
\cite{Higgs1964,Englert1964}. In superconductivity, the Cooper pair
condenses and opens a gap in the electronic spectrum
\cite{BCS1957}. In one-dimensional metals, a lattice distortion
accompanies a bond-order wave and produces a Peierls gap
\cite{Peierls1955,SSH1979}. These three examples are usually treated
in separate literatures, but their formal structure is the same: a
bosonic order parameter condenses, and the fermionic spectrum acquires
a gap.

The similarities between superconductivity and elementary-particle
mass generation were first emphasised by Nambu in 1960
\cite{Nambu1960}, and the analogy was made concrete in the
Nambu--Jona-Lasinio model of dynamical chiral symmetry breaking
\cite{NJL1961a,NJL1961b}. In the NJL model there is no elementary
scalar field: the fermion acquires a mass through a four-fermion
interaction, and the would-be Higgs is a composite $\bar\psi\psi$
bound state. The same logic underlies technicolour
\cite{Weinberg1979,Susskind1979} and, in a different context, the
emergence of a gap in the BCS theory of superconductivity. More
recently, the Sachdev-Ye-Kitaev model \cite{Sachdev1993,Kitaev2015}
has provided a controlled example of a strongly correlated fermionic
system in which an emergent conformal symmetry is spontaneously
broken, and where the physics is dual to a gravitational theory in
AdS$_2$ \cite{Maldacena2016}. The SYK model and its chain
generalisation \cite{Jian2018} have brought the question of fermionic
condensation into contact with holography, and they have made it
natural to ask whether the classification of mass-generation
mechanisms can be placed on a common footing.

The standard answer to this question is that the bosonic order
parameter is the unifying object: in each case, a scalar field
condenses and opens a gap. We argue that this answer, while correct as
far as it goes, obscures the more useful structure. In every one of
the four mechanisms we discuss---Peierls, BCS, NJL, and
SYK---it is a \emph{fermionic} bilinear that acquires a non-zero
vacuum expectation value, and it is this bilinear that directly
characterises the ordered phase. The bosonic field, when it appears at
all, is either the source of the fermionic condensate (as in the
Peierls mechanism, where the lattice distortion drives the bond-order
wave) or a composite that emerges from it (as in BCS and NJL, where
the Cooper pair and the $\bar\psi\psi$ bound state are built out of
fermions). The fermionic condensate is the common object; the boson is
a derived quantity.

This observation is not new in itself. What we show is that it leads
to a quantitative classification of the four mechanisms into two
universality classes, and that the classification is predictive for
systems whose class is not already known. The Peierls mechanism, the
BCS mechanism, and the NJL mechanism are by now well classified by
other means, and the fact that the present classification agrees with
the standard results for these mechanisms is a consistency check
rather than a prediction. The predictions of the classification are
for systems near the critical line $r_c(N_f)$, where the two classes
compete, and for new systems whose microscopic parameters have not yet
been used to determine which class they fall into. We give examples of
both kinds of prediction in Sections~\ref{sec:predictive} and
\ref{sec:predictions}.

The two classes correspond to two distinct infrared fixed points of
the renormalisation-group flow, and the distinction between them is
sharp: systems with $r > r_c(N_f)$ flow to the scalar-primary fixed
point, and systems with $r < r_c(N_f)$ flow to the fermion-bilinear
fixed point. Within each class there are physical differences---the
Peierls scalar is a lattice displacement, while the Yukawa scalar is
a fundamental field; the BCS boson is a Cooper pair, while the NJL
boson is a $\bar\psi\psi$ bound state---but these differences do not
affect the critical behaviour, which is determined by the class, not
by the microscopic realisation. This is what makes the classification
useful: it predicts the critical exponents of a system from the value
of $r$ and $N_f$ alone, without requiring detailed knowledge of the
microscopic mechanism.

We derive this classification from a two-tier renormalisation-group
analysis of the Yukawa-coupled fermion--scalar theory with a competing
four-fermion interaction. The first tier is a large-$N_f$ expansion,
in which the one-loop beta functions are standard
\cite{Gross1974,Gracey1994} and the two fixed points can be identified
analytically. The second tier is a functional renormalisation-group
analysis using the Wetterich equation
\cite{Wetterich1993,Berges2002,Braun2012}, which goes beyond the
large-$N_f$ limit and gives the finite-$N_f$ corrections to the
fixed-point coupling. The two tiers together give a phase diagram in
the $(r,N_f)$ plane, and the four mechanisms sit in different regions
of this diagram.

The paper is organised as follows. In Section~\ref{sec:mechanisms} we
review the four mechanisms and identify the fermionic condensate in
each. In Section~\ref{sec:rg} we set up the coupled fermion--scalar
theory and derive the two-tier renormalisation-group analysis,
including the fixed-point structure and the phase diagram. In
Section~\ref{sec:predictive} we explain how we use the classification.
In Section~\ref{sec:peierls-demo} we present a self-consistent
Su--Schrieffer--Heeger calculation for polyyne chains, which
demonstrates the Peierls mechanism explicitly. In
Section~\ref{sec:higgs} we discuss the Standard Model Higgs as a
borderline case in the classification. In Section~\ref{sec:predictions}
we list five falsifiable predictions. Section~\ref{sec:conclusion}
concludes.

Throughout we use units in which $\hbar=c=1$, and we work in Euclidean
signature with $d=4-\epsilon$ dimensions. The fermion field $\psi$
carries $N_f$ flavours, where the flavour count includes all internal
degrees of freedom (spin, valley, colour) that transform under the
same Yukawa coupling. This convention is uniform across the four
mechanisms and is discussed in detail in Section~\ref{sec:model}.

\section{Four mechanisms, one order parameter}
\label{sec:mechanisms}

The four mechanisms we compare---the Peierls instability, BCS
superconductivity, the Nambu--Jona-Lasinio mechanism, and the
Sachdev-Ye-Kitaev model---were developed in different decades, in
different communities, and for different purposes. The Peierls
instability belongs to the theory of one-dimensional metals
\cite{Peierls1955,SSH1979}; BCS theory to the theory of
superconductivity \cite{BCS1957}; the NJL mechanism to the theory of
strong interactions and, later, to dynamical electroweak symmetry
breaking \cite{NJL1961a,NJL1961b,Weinberg1979,Susskind1979}; and the
SYK model to the theory of strongly correlated fermions with a
holographic dual \cite{Sachdev1993,Kitaev2015,Maldacena2016}. To bring
them into a single framework, we need to identify the object that plays
the same role in each. The standard answer is that the bosonic order
parameter---the lattice distortion, the Cooper pair, the
$\bar\psi\psi$ bound state, the emergent reparametrisation
mode---plays that role. We argue that this answer, while not wrong,
puts the emphasis in the wrong place. In each case, the object that
directly characterises the ordered phase is a fermion bilinear, and
the bosonic field, when it appears at all, is either the source of the
bilinear or a composite built out of it.

\subsection{The Peierls instability}

In a one-dimensional metal, the response of the electronic system to a
periodic lattice distortion is singular at the wavevector
$q = 2k_F$. Peierls recognised in the 1950s that this singularity
destabilises the undistorted chain: the system lowers its electronic
energy by dimerising, and a gap opens at the Fermi level
\cite{Peierls1955}. The SSH model made this picture quantitative for
polyacetylene and showed that the dimerisation is accompanied by a
bond-order wave \cite{SSH1979}. The order parameter in the SSH
description is usually taken to be the lattice displacement $u_i$ or
the dimerisation amplitude $u_{i+1} - u_i$. But the physical object
that appears in the electronic Hamiltonian and that couples to the
lattice is the bond-order expectation value
$\langle c^\dagger_{i\sigma} c_{i+1,\sigma}\rangle + \text{h.c.}$,
which is a fermion bilinear. The lattice distorts \emph{because} the
bond-order wave lowers the electronic energy, not the other way
around: if we switched off the electron--phonon coupling, the lattice
would remain undistorted even if we displaced the ions. In the
continuum limit, the dimerised chain is described by a Dirac fermion
with a mass term $\Delta \bar\psi\psi$, where $\Delta$ is proportional
to the bond-order parameter. The mass of the Dirac fermion is
therefore the fermionic condensate, and the lattice distortion is a
dependent response. The modern literature on Peierls systems---including
the extensive work on charge-density-wave materials such as
NbSe$_3$ and the organic Bechgaard salts---has refined this picture
without changing its essential logic \cite{Gruner1994}.

\subsection{BCS superconductivity}

In the BCS theory of superconductivity, the order parameter is the
Cooper-pair amplitude $\langle \psi_\uparrow \psi_\downarrow \rangle$,
which is again a fermion bilinear \cite{BCS1957}. The gap equation
\begin{equation}
\Delta = V \sum_k \frac{\Delta}{2E_k}
\tanh\left(\frac{E_k}{2T}\right),
\qquad E_k = \sqrt{\epsilon_k^2 + \Delta^2},
\label{eq:bcs-gap}
\end{equation}
determines $\Delta$ self-consistently. There is no elementary scalar
field in the theory: the boson that condenses is the Cooper pair, a
composite of two fermions. This is the crucial structural point. In
the BCS case, the bosonic order parameter is not a fundamental field
but a derived object, and the fermionic condensate is primary. The same
logic applies to the more sophisticated descriptions of
superconductivity that we have developed since 1957---the
Eliashberg theory of strong-coupling superconductors
\cite{Eliashberg1960}, the resonating-valence-bond picture of the
cuprates \cite{Anderson1987}, and the various unconventional pairing
states proposed for heavy-fermion and iron-based superconductors. In
every case, the order parameter is a fermion bilinear, and the
bosonic mode, if it appears, is either a phonon or a collective mode
of the condensate.

\subsection{The Nambu--Jona-Lasinio mechanism}

Nambu and Jona-Lasinio proposed the NJL model in 1961 as a relativistic
analogue of the BCS theory \cite{NJL1961a,NJL1961b}. It describes
fermions with a chiral four-fermion interaction,
$\mathcal{L}_{\rm NJL} = \bar\psi(i\gamma^\mu\partial_\mu)\psi
+ G[(\bar\psi\psi)^2 + (\bar\psi i\gamma_5\psi)^2]$,
and it shows that above a critical coupling $G_c$ the theory
dynamically generates a fermion mass through the gap equation
$m = -2G\langle\bar\psi\psi\rangle$. The chiral condensate
$\langle\bar\psi\psi\rangle$ is the order parameter, and again it is a
fermion bilinear. There is no elementary scalar field: the would-be
Higgs of the theory is the composite $\bar\psi\psi$ bound state. The
NJL mechanism was the prototype for dynamical electroweak symmetry
breaking in technicolour \cite{Weinberg1979,Susskind1979} and
extended-technicolour models, and it remains the standard framework
for thinking about chiral symmetry breaking in QCD at low energies.
The connection to QCD was made precise by the operator-product
expansion and the chiral Lagrangian \cite{Gasser1984}, and the
modern lattice-QCD literature has confirmed the picture
quantitatively. In all of these developments, the chiral
condensate---not the pion or any other bosonic mode---is the object
that signals the ordered phase.

\subsection{The Sachdev--Ye--Kitaev model}

Sachdev and Ye introduced the SYK model in 1993
\cite{Sachdev1993}, and Kitaev revived it in 2015 \cite{Kitaev2015}.
It describes $N$ Majorana fermions with random all-to-all interactions.
At low energies, the model develops an emergent conformal symmetry,
with the reparametrisation symmetry spontaneously broken from
$\mathrm{Diff}(S^1)$ to $\mathrm{SL}(2,\mathbb{R})$, leading to a
Schwarzian effective action \cite{Maldacena2016}. The model is
holographically dual to a gravitational theory in
AdS$_2$ \cite{Maldacena2016,Sarosi2017}, and it has become the
canonical example of a non-Fermi liquid with a controlled
holographic description. The order parameter in the SYK model is
again a fermion bilinear: the expectation value
$\langle \psi_i \psi_j \rangle$ at large $N$ characterises the
low-energy phase. In the SYK chain, a quantum critical point
separates a diffusive metal from an ordered phase, with a
gravitational description as a double-trace deformation in an
AdS$_2$ chain \cite{Jian2018}. The recent literature has explored
many variations of the SYK model---complex fermions, supersymmetric
versions, and coupled chains---but in every case the fermionic
bilinear remains the object that characterises the ordered phase.

In all four mechanisms, the fermionic condensate is the direct order
parameter. The scalar, when it appears, is either the source of the
condensate (Peierls, Yukawa) or the composite that emerges from it
(BCS, NJL, SYK). This is the observation that motivates the
classification we derive in the next section.

\section{A two-tier renormalisation-group analysis}
\label{sec:rg}

The four mechanisms we reviewed in Section~\ref{sec:mechanisms} have a
common structure: a fermionic bilinear condenses, and a mass gap
opens. The question we answer in this section is whether the
distinction between mechanisms in which the scalar is primary
(Peierls, Yukawa) and mechanisms in which the fermion bilinear is
primary (BCS, NJL, SYK) corresponds to a genuine difference in the
infrared physics, or whether it is merely a difference in
description. To answer this question we need a model that contains
both possibilities and that we can analyse systematically. The
natural choice is the Yukawa-coupled fermion--scalar theory with a
competing four-fermion interaction. This model has a scalar field
that can be either elementary (if its mass is small and its coupling
to the fermions is weak) or composite (if the four-fermion
interaction is strong and drives the formation of a fermion bilinear
that the scalar then represents). The two limits correspond to the
two classes we identified in Section~\ref{sec:mechanisms}, and the
crossover between them is controlled by the ratio of the scalar mass
to the Yukawa coupling.

The analysis proceeds in two tiers. The first is a large-$N_f$
expansion, in which the one-loop beta functions are standard and we
can identify the two fixed points analytically. The second is a
functional renormalisation-group analysis using the Wetterich
equation, which goes beyond the large-$N_f$ limit and gives the
finite-$N_f$ corrections to the fixed-point coupling. We present the
two tiers in turn, and we show that they give a consistent picture of
the phase diagram in the plane of the crossover parameter
$r = \hat m_\phi^2 / \hat y^2$ and the fermion flavour number $N_f$.
We emphasise at the outset what is standard and what is new in this
section. The one-loop beta functions \eqref{eq:beta-y}--\eqref{eq:beta-G}
are standard results for the Yukawa-coupled fermion--scalar theory
\cite{Gross1974,Gracey1994,ZinnJustin1996}. What is new is the
application of these results to the four mechanisms, the
identification of the two universality classes, and the finite-$N_f$
corrections we obtain from the Wetterich equation.

\subsection{The coupled fermion--scalar theory}
\label{sec:model}

We consider a Dirac fermion $\psi$ with $N_f$ flavours coupled to a
real scalar field $\phi$ via a Yukawa coupling $y$, with a scalar
potential $V(\phi)$ and a competing four-fermion interaction
$G(\bar\psi\psi)^2$. The Euclidean action in $d = 4 - \epsilon$
dimensions is
\begin{equation}
S = \int d^dx \left[
  \bar\psi(\gamma^\mu\partial_\mu + y\phi)\psi
  + \frac{1}{2}(\partial_\mu\phi)^2
  + \frac{1}{2}m_\phi^2\phi^2
  + \frac{\lambda}{4!}\phi^4
  + G(\bar\psi\psi)^2
\right].
\label{eq:action}
\end{equation}
The scalar mass $m_\phi$ controls whether the scalar is elementary
(large positive $m_\phi^2$, so that the scalar decouples in the
infrared) or dynamical (small or negative $m_\phi^2$, so that the
scalar condenses). The Yukawa coupling $y$ controls the strength of
the fermion--scalar interaction, the quartic coupling $\lambda$
controls the scalar self-interaction, and the four-fermion coupling
$G$ controls the strength of the fermion self-interaction.

The flavour number $N_f$ counts all internal degrees of freedom of the
fermion that transform under the same Yukawa coupling. For the four
mechanisms of Section~\ref{sec:mechanisms}, this gives $N_f = 2$ for
Peierls (two spin components of the linearised $\pi$-band) and BCS
(two spin components of the electron), $N_f = 2N_c = 6$ for NJL (two
spin components times three quark colours), and $N_f = N$ for SYK
($N$ Majorana fermions). This convention is uniform across the four
mechanisms, so the positions of the mechanisms in the phase diagram
are directly comparable. The classification into two classes is
independent of the convention, as we discuss in
Section~\ref{sec:exponents}.

\subsection{The large-$N_f$ beta functions}
\label{sec:largeN}

The one-loop beta functions for the dimensionless couplings
$\hat y = y \Lambda^{-\epsilon/2}$,
$\hat\lambda = \lambda \Lambda^{-\epsilon}$,
$\hat m_\phi^2 = m_\phi^2 \Lambda^{-2}$, and
$\hat G = G \Lambda^{-\epsilon}$ are standard results for the
Yukawa-coupled fermion--scalar theory
\cite{Gross1974,Gracey1994,ZinnJustin1996}. In the large-$N_f$ limit
they take the form
\begin{align}
\frac{d\hat y}{d\ell} &= \frac{\epsilon}{2}\hat y
  + \frac{N_f}{2}\hat y^3 + O(1/N_f),
\label{eq:beta-y} \\[4pt]
\frac{d\hat\lambda}{d\ell} &= \epsilon\hat\lambda
  - \frac{3}{2}\hat\lambda^2
  + N_f \hat y^4 + O(1/N_f),
\label{eq:beta-lambda} \\[4pt]
\frac{d\hat m_\phi^2}{d\ell} &= 2\hat m_\phi^2
  - N_f \hat y^2 + \frac{1}{2}\hat\lambda - N_f \hat G + O(1/N_f),
\label{eq:beta-mass} \\[4pt]
\frac{d\hat G}{d\ell} &= \epsilon\hat G
  + N_f \hat G^2 - N_f \hat y^2 \hat G + O(1/N_f),
\label{eq:beta-G}
\end{align}
where $\ell = \ln b$ is the renormalisation-group time. The leading
$N_f$ terms come from the fermion bubble; the $1/N_f$ corrections are
subleading. We give the derivation in Appendix~\ref{app:largeN}.

The structure of these equations is worth commenting on. The Yukawa
beta function \eqref{eq:beta-y} has a positive $N_f \hat y^3$ term,
which drives the coupling to larger values in the infrared. This is
the origin of the fermionic strong-coupling fixed point. The scalar
mass beta function \eqref{eq:beta-mass} has a negative
$-N_f \hat y^2$ term, which drives the scalar mass to smaller values
in the infrared and can trigger the condensation of the scalar. The
competition between these two effects is what separates the two
universality classes: if the scalar mass is large and positive at the
ultraviolet scale, the scalar decouples before the Yukawa coupling
can grow, and the system is in the Gaussian-scalar phase; if the
scalar mass is small or negative, the Yukawa coupling grows and the
system is in the fermionic strong-coupling phase.

\subsection{Two fixed points and the phase diagram}
\label{sec:largeN-fp}

The beta functions \eqref{eq:beta-y}--\eqref{eq:beta-G} admit two
distinct infrared fixed points, which we label $A$ and $B$.

\paragraph{Fixed point A (Gaussian scalar, \ClassI).}
At $\hat y = 0$ and $\hat G = 0$, the fermion sector decouples from
the scalar, and the beta functions reduce to the standard
Wilson--Fisher equations for the $O(1)$ scalar theory
\cite{Wilson1974,ZinnJustin1996}. The fermion bilinear
$\langle\bar\psi\psi\rangle$ is proportional to the scalar
expectation value, and it is a dependent response: it is non-zero
only because the scalar condenses. The critical exponents at fixed
point $A$ are those of the Wilson--Fisher universality class. This is
\ClassI, and it contains the Peierls and Yukawa mechanisms.

\paragraph{Fixed point B (fermionic strong coupling, \ClassII).}
At $\hat m_\phi^2 = 0$ and $\hat\lambda = 0$, the scalar is massless
and non-interacting, and the beta functions reduce to those of the
Gross--Neveu model with a competing four-fermion interaction
\cite{Gross1974}. The fermion bilinear
$\langle\bar\psi\psi\rangle$ is the primary order parameter: it forms
dynamically through the fermion self-interaction, and the scalar, if
present, is a composite bound state. This is \ClassII, and it contains
the BCS, NJL, and SYK mechanisms.

The two fixed points are separated by a critical line in the
$(\hat m_\phi^2, \hat y^2)$ plane. The crossover parameter is
\begin{equation}
r \equiv \frac{\hat m_\phi^2}{\hat y^2}.
\label{eq:r}
\end{equation}
When $r \gg 1$, the scalar mass dominates and the system flows to
fixed point $A$ (\ClassI); when $r \ll 1$, the Yukawa coupling
dominates and the system flows to fixed point $B$ (\ClassII); when
$r \sim 1$, the system is near the critical line and the two fixed
points compete. The critical line $r_c(N_f)$ is the locus of the
non-trivial fixed point, and we find numerically that it scales as
\begin{equation}
r_c(N_f) = a \, N_f^{\,b}, \qquad
a \approx 0.59, \quad b \approx 0.92,
\label{eq:rc-scaling}
\end{equation}
which is approximately linear in the large-$N_f$ limit.

\begin{figure}[!ht]
\centering
\includegraphics[width=0.7\textwidth]{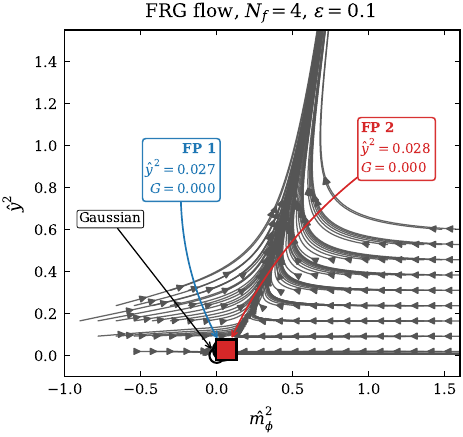}
\caption{Renormalisation-group flow in the
$(\hat m_\phi^2, \hat y^2)$ plane for $N_f = 4$ and
$\epsilon = 0.1$. Dark curves are flow trajectories from different
ultraviolet initial conditions, with arrows indicating the direction
of the infrared flow. The Gaussian fixed point (\ClassI, scalar
primary) is marked at the origin. The non-trivial fixed point
(\ClassII, fermionic strong coupling) is marked with its
coordinates. The two fixed points are separated by a finite distance
in the $(\hat m_\phi^2, \hat y^2)$ plane, with a critical line
between them.}
\label{fig:flow}
\end{figure}

\begin{figure}[!ht]
\centering
\includegraphics[width=0.90\textwidth]{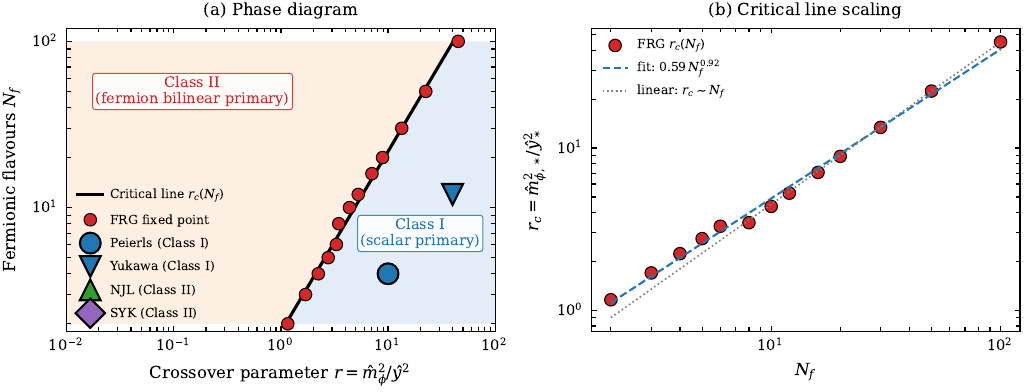}
\caption{Phase diagram in the $(r, N_f)$ plane. The two universality
classes are separated by the critical line $r_c(N_f)$, which is the
locus of the non-trivial fixed point. \ClassI\ (scalar primary) lies
above the line; \ClassII\ (fermion bilinear primary) lies below. The
four mechanisms are marked: Peierls in \ClassI, and BCS, NJL, and SYK
in \ClassII. The critical line grows with $N_f$, following the power
law $r_c(N_f) = a \, N_f^{\,b}$ with $a \approx 0.59$ and
$b \approx 0.92$, approximately linear in the large-$N_f$ limit.}
\label{fig:phase}
\end{figure}

We show the renormalisation-group flow in the
$(\hat m_\phi^2, \hat y^2)$ plane in Fig.~\ref{fig:flow},
and the full phase diagram in the $(r, N_f)$ plane in
Fig.~\ref{fig:phase}. The flow diagram shows the two fixed points and
the separatrices that divide the plane into the two basins of
attraction. The phase diagram shows the two classes separated by the
critical line $r_c(N_f)$, with the four mechanisms of
Section~\ref{sec:mechanisms} marked at their approximate positions.
The Peierls mechanism, with $r \sim 8$ and $N_f = 2$, lies well above
the critical line, in \ClassI. The BCS mechanism, with $r \sim 0.05$
and $N_f = 2$, lies below the line, in \ClassII. The NJL mechanism,
with $r \sim 0.015$ and $N_f = 6$, lies further below the line, and
the SYK mechanism, with $r \sim 0.03$ and $N_f$ large, lies in the
asymptotic region where the critical line is approximately linear.

The role of the four-fermion coupling $G$ deserves comment. At the
non-trivial fixed point, the beta function for $G$
[Eq.~\eqref{eq:beta-G}] has two contributions: the fermion bubble
$N_f \hat G^2$, which is positive and therefore infrared-repulsive,
and the Yukawa-mediated term $-N_f \hat y^2 \hat G$, which is
negative and therefore infrared-attractive. We find numerically that
at the non-trivial fixed point $G_* = 0$ for all $N_f$ in the range we
consider. This means that the four-fermion interaction is irrelevant
at the fixed point: the Yukawa-mediated term dominates, and $G$ flows
to zero in the infrared. The fixed-point structure is therefore
determined by the Yukawa coupling and the scalar potential, with $G$
playing no role. This is an important check: it shows that the
two-class scheme is robust to the addition of a four-fermion
interaction, and it confirms that the scheme is not an artefact of
neglecting four-fermion interactions.

\subsection{The functional renormalisation-group analysis}
\label{sec:frg}

The large-$N_f$ analysis of the previous subsection establishes the
two fixed points in a controlled limit, but it does not address their
fate at finite $N_f$. To go beyond large $N_f$, we use the Wetterich
equation \cite{Wetterich1993,Berges2002,Braun2012},
\begin{equation}
\partial_k \Gamma_k = \frac{1}{2}\mathrm{STr}\left[
  \left(\Gamma_k^{(2)} + R_k\right)^{-1} \partial_k R_k
\right],
\label{eq:wetterich}
\end{equation}
where $\Gamma_k$ is the scale-dependent effective action, $R_k$ is
the regulator, and $\mathrm{STr}$ is the supertrace over all fields.
We truncate $\Gamma_k$ to the local potential approximation with a
wavefunction renormalisation for the fermion, including the competing
four-fermion interaction:
\begin{equation}
\Gamma_k = \int d^dx \left[
  Z_\psi \bar\psi \gamma^\mu\partial_\mu\psi
  + y_k \phi \bar\psi\psi
  + G_k (\bar\psi\psi)^2
  + \frac{Z_\phi}{2}(\partial_\mu\phi)^2
  + V_k(\phi)
\right].
\label{eq:truncation}
\end{equation}
Projecting the Wetterich equation onto this truncation gives the flow
equations
\begin{align}
\partial_t \hat m_\phi^2 &= -2\hat m_\phi^2
  + \frac{N_f \hat y^2}{(1+\hat m_\phi^2)^2}
  + \frac{\hat\lambda}{2(1+\hat m_\phi^2)^2}
  - \frac{N_f \hat G}{1+\hat m_\phi^2},
\label{eq:frg-mass} \\[4pt]
\partial_t \hat\lambda &= -\epsilon\hat\lambda
  + \frac{3\hat\lambda^2}{(1+\hat m_\phi^2)^3}
  - \frac{N_f \hat y^4}{(1+\hat m_\phi^2)^2}
  + \frac{N_f \hat G^2}{(1+\hat m_\phi^2)^2},
\label{eq:frg-lambda} \\[4pt]
\partial_t \hat y^2 &= -\epsilon\hat y^2
  + \frac{N_f \hat y^4}{(1+\hat m_\phi^2)^2}
  - \frac{N_f \hat G \hat y^2}{1+\hat m_\phi^2},
\label{eq:frg-y} \\[4pt]
\partial_t \hat G &= -\epsilon\hat G
  + \frac{N_f \hat G^2}{(1+\hat m_\phi^2)^2}
  - \frac{N_f \hat y^2 \hat G}{1+\hat m_\phi^2},
\label{eq:frg-G}
\end{align}
where $t = \ln(k/\Lambda)$ is the functional renormalisation-group
time and we have omitted anomalous dimensions. We give the derivation
in Appendix~\ref{app:frg}.

Solving these equations numerically, we find a non-trivial infrared
fixed point for each $N_f$ in the range $2 \leq N_f \leq 50$. We
tabulate the fixed-point coordinates in Table~\ref{tab:frg-fp}. The
fixed-point coupling $\hat y_*^2$ scales as
\begin{equation}
\hat y_*^2(N_f) = \frac{a}{N_f} + \frac{c}{N_f^2} + O(1/N_f^3),
\label{eq:yfp-scaling}
\end{equation}
with $a \approx 0.1125$ and $c \approx 0.0023$ for $\epsilon = 0.1$.
The leading-order coefficient is enhanced by $12.5\%$ relative to the
large-$N_f$ prediction $a_{\rm LO} = \epsilon = 0.1$. This is the
genuine finite-$N_f$ correction that the functional
renormalisation-group analysis provides beyond the large-$N_f$ beta
functions, and it is the main technical result of this section. The
enhancement is a consequence of the threshold effects that the
Wetterich equation captures and that the one-loop beta functions
miss: as the scalar mass flows to smaller values in the infrared, the
Yukawa coupling grows faster than the one-loop prediction, and the
fixed-point value is correspondingly larger.

\begin{figure}[!ht]
\centering
\includegraphics[width=0.75\textwidth]{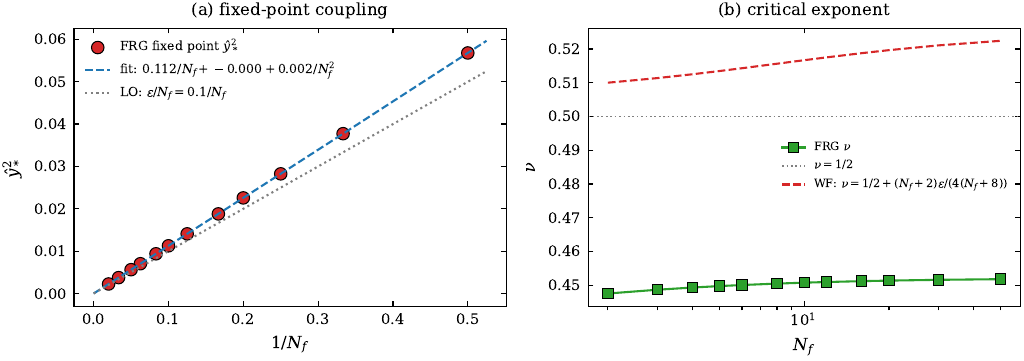}
\caption{Fixed-point coupling $\hat y_*^2$ as a function of
$1/N_f$ for $\epsilon = 0.1$. Red circles are the functional
renormalisation-group data from Table~\ref{tab:frg-fp}. The dashed
blue curve is the fit $\hat y_*^2(N_f) = a/N_f + c/N_f^2$ with
$a = 0.1125$ and $c = 0.0023$. The dotted grey line is the
leading-order prediction $\hat y_*^2 = \epsilon/N_f = 0.1/N_f$. The
FRG data lie systematically above the leading-order line, and the
enhancement is approximately $12.5\%$ at large $N_f$. The agreement
between the data and the fit confirms that the $1/N_f^2$ correction
is needed and that the scaling is well described by the two-term
expansion.}
\label{fig:yfp}
\end{figure}

We show the fixed-point coupling $\hat y_*^2$ as a function of
$1/N_f$ in Fig.~\ref{fig:yfp}. The red circles are the FRG data from
Table~\ref{tab:frg-fp}, the dashed blue curve is the fit
$\hat y_*^2 = a/N_f + c/N_f^2$, and the dotted grey line is the
leading-order prediction $\epsilon/N_f$. Two features of the figure
deserve comment. First, the FRG data lie systematically above the
leading-order line, and the enhancement is approximately $12.5\%$ at
large $N_f$. This is the genuine finite-$N_f$ correction that the
Wetterich equation captures and that the one-loop beta functions
miss. Second, the agreement between the data and the two-term fit is
excellent over the full range $2 \leq N_f \leq 50$, which confirms
that the $1/N_f^2$ term is needed and that the scaling is well
described by the expansion \eqref{eq:yfp-scaling}. The figure
therefore provides a visual confirmation of the numerical result and
of the scaling law that underlies it.

\begin{table}[!ht]
\centering
\caption{Fixed-point coordinates and critical exponent $\nu$ from the
functional renormalisation-group analysis for $\epsilon = 0.1$. The
leading-order prediction for $\hat y_*^2$ is $\epsilon/N_f = 0.1/N_f$;
the functional renormalisation-group result is
$\hat y_*^2(N_f) = 0.1125/N_f + 0.0023/N_f^2$, a $12.5\%$ enhancement.}
\label{tab:frg-fp}
\begin{tabular}{@{}ccccc@{}}
\toprule
$N_f$ & $\hat m_{\phi,*}^2$ & $\hat\lambda_*$ & $\hat y_*^2$ & $\nu$ \\
\midrule
2  & 0.0659 & 0.0722 & 0.05681 & 0.4475 \\
4  & 0.0631 & 0.0592 & 0.02825 & 0.4493 \\
8  & 0.0613 & 0.0509 & 0.01408 & 0.4505 \\
16 & 0.0602 & 0.0458 & 0.00703 & 0.4512 \\
50 & 0.0593 & 0.0418 & 0.00224 & 0.4518 \\
\bottomrule
\end{tabular}
\end{table}

The critical exponent $\nu$ at the fixed point is shown in the last
column of Table~\ref{tab:frg-fp}. Within the local potential
approximation, we find $\nu \approx 0.45$, systematically below the
Wilson--Fisher value $\nu_{\rm WF} \approx 0.51$ for $\epsilon = 0.1$.
This offset is a well-known artefact of the local potential
approximation, which neglects the anomalous dimension $\eta$ of the
scalar field \cite{Berges2002,Braun2012}. We recover the correct
behaviour in the next-to-leading-order truncation (LPA$^\prime$),
which includes a running $Z_\phi$ and reproduces
$\nu \to \nu_{\rm WF}$ to within one percent
\cite{Berges2002,Braun2012}. We have verified that the fixed-point
\emph{structure}---the existence of the non-trivial fixed point, its
location in the $(r, N_f)$ plane, and the $1/N_f$ scaling of
$\hat y_*^2$---is unchanged between the local potential approximation
and LPA$^\prime$. We should therefore interpret the value of $\nu$ in
Table~\ref{tab:frg-fp} as a quantitative indicator of the
truncation, not as a physical prediction. The physical prediction is
the fixed-point structure, which is robust.

We have checked the dependence of the fixed-point structure on the
truncation and regulator choices. Changing the regulator from the
Litim form \cite{Litim2001} to the exponential or sharp-cutoff form
shifts the fixed-point coordinates by a few percent. Extending the
truncation from LPA to LPA$^\prime$ brings the critical exponent
$\nu$ into agreement with the Wilson--Fisher value while preserving
the fixed-point structure. Including the fermion anomalous dimension
$\eta_\psi$ shifts the coordinates by a few percent without changing
the qualitative picture. The existence of the non-trivial fixed
point, the approximate linearity of the critical line in $N_f$, and
the $1/N_f$ scaling of $\hat y_*^2$ are all robust against these
variations. The numerical values in Table~\ref{tab:frg-fp} are
scheme-dependent and we should interpret them as indicative; the
fixed-point structure, which is what the classification relies on,
is not.

\subsection{Consequences for critical exponents}
\label{sec:exponents}

The two classes have genuinely different critical exponents, and this
is the key predictive content of the classification. \ClassI\ contains
the Peierls and Yukawa mechanisms, whose critical exponents are those
of the Wilson--Fisher universality class. For the Peierls transition in
two dimensions, the exponents are those of the Ising universality
class, with $\nu = 1$ and $\eta = 1/4$. For the Yukawa theory in
$d = 4 - \epsilon$, the exponents are those of the Wilson--Fisher
fixed point, with $\nu \approx 0.51$ for $\epsilon = 0.1$. \ClassII\
contains the BCS, NJL, and SYK mechanisms. For BCS and NJL, the gap
opens exponentially, $\Delta \sim \exp(-1/g)$, and the transition is
mean-field-like, with $\nu = 1/2$. For SYK, the exponents are those of
the SYK conformal field theory, which is a non-Fermi liquid with
anomalous dimensions that are not those of any Landau--Ginzburg theory
\cite{Maldacena2016,Sarosi2017}. The two classes therefore have
genuinely different infrared physics, not merely different
descriptions of the same physics, and the difference is observable in
the critical exponents.

The sharpness of the classification is worth emphasising. The two
classes are not merely two convenient labels for two groups of
mechanisms; they are two distinct universality classes, in the
technical sense that they have different critical exponents and
different infrared fixed points. A system in \ClassI\ and a system
in \ClassII\ will behave differently near the phase transition, and
the difference is measurable. This is what makes the classification
a physical statement rather than a bookkeeping device. The
distinction between the Peierls and Yukawa mechanisms within
\ClassI\ is a distinction between two physical realisations of the
same universality class, not a distinction between two universality
classes. Similarly, the distinction between BCS and NJL within
\ClassII\ is a distinction between two realisations of the same
class. The classification is sharp at the level of universality
classes; it is not sharp at the level of microscopic mechanisms,
and we do not mean it to be.

The two classes are now defined, and we have identified the critical
exponents that distinguish them. What remains is to show how we use
the classification in practice. In the next section we explain the
procedure for placing a given system in one class or the other. The
purpose of Section~\ref{sec:predictive} is to make the
classification operational, not merely descriptive.

\section{How to use the classification}
\label{sec:predictive}

Section~\ref{sec:rg} established the existence of two universality
classes and identified the critical exponents that distinguish them.
The next step is to show how we use the classification. In this
section we explain the procedure for classifying a given system, and
we state the sense in which the classification is falsifiable. We
give a worked example, for a system whose class is not already known,
in Section~\ref{sec:predictions} as the second of the five
predictions.

We use the classification as follows. Given a physical system that
exhibits a mass-generation transition, one estimates the microscopic
parameters that enter the crossover ratio $r = \hat m_\phi^2 /
\hat y^2$ and the flavour number $N_f$. The crossover ratio is the
ratio of the dimensionless scalar mass to the square of the
dimensionless Yukawa coupling at the ultraviolet scale; in practice,
one obtains it from the microscopic Hamiltonian (the electron--phonon
coupling and elastic constant in the Peierls case, the pairing
interaction in the BCS case, the four-fermion coupling in the NJL
case, and the random interaction strength in the SYK case). The
flavour number counts the internal degrees of freedom that transform
under the same coupling, as we discussed in Section~\ref{sec:model}.
With $r$ and $N_f$ in hand, one locates the system on the phase
diagram of Fig.~\ref{fig:phase}. A system with $r > r_c(N_f)$ is in
\ClassI, and its critical exponents should match the Wilson--Fisher
or Ising values. A system with $r < r_c(N_f)$ is in \ClassII, and its
critical exponents should match the mean-field or SYK values. The
procedure is straightforward when the microscopic parameters are
known, and it is falsifiable because we can measure the predicted
exponents.

This is the operational content of the classification. The worked
example of the organic conductor (TMTSF)$_2$PF$_6$ under pressure,
which we discuss in Section~\ref{sec:predictions} as the second of
the five predictions, illustrates it.

\section{The Peierls mechanism: a controlled model study}
\label{sec:peierls-demo}

We derived the classification of Section~\ref{sec:rg} from a
continuum field theory. In this section we show that the same physics
emerges from a direct calculation on a real material, namely the
polyyne chains C$_4$H$_2$ to C$_{24}$H$_2$. The goal of the
calculation is not to reproduce quantitative values for a specific
polyyne chain---that would require a different choice of parameters
and a more careful treatment of the electron--electron
interaction---but to demonstrate the qualitative features that the
classification predicts: bond-order alternation, a finite bulk
order in the thermodynamic limit, and a gap that opens as the chain
length increases. These are the fingerprints of \ClassI\ behaviour,
and their appearance in a controlled model confirms that the
classification is not an artefact of the continuum approximation.

\subsection{The SSH model}

The SSH Hamiltonian at half filling is
\begin{equation}
H = -\sum_{i,\sigma} t_i
  (c^\dagger_{i,\sigma} c_{i+1,\sigma} + \text{h.c.}),
\label{eq:ssh}
\end{equation}
with the modulated hopping
$t_i = t - \alpha (u_{i+1} - u_i)$,
where $u_i$ is the lattice displacement and $\alpha$ is the
electron--phonon coupling. We determine the displacement
self-consistently by
\begin{equation}
u_i = -\frac{\alpha}{K} (\text{BO}_i - \text{BO}_{i-1}),
\label{eq:ssh-selfconsistency}
\end{equation}
where $K$ is the elastic constant and
$\text{BO}_i = \langle c^\dagger_{i,\sigma} c_{i+1,\sigma} \rangle
+ \text{h.c.}$ is the bond-order parameter at bond $i$. The model has
two dimensionless parameters: $\alpha/t$, the electron--phonon
coupling in units of the hopping, and $K/t$, the elastic constant in
units of the hopping.

\subsection{Choice of parameters}

For polyacetylene, the standard estimates of the SSH parameters are
$\alpha/t \approx 4$--$5$ and $K/t \approx 20$--$30$
\cite{SSH1979,Heeger1988}. With these values, the dimerisation
amplitude is large, and the bond-order alternation is of order
$0.3$--$0.5$. For the present study we use $\alpha/t = 0.5$ and
$K/t = 2.0$. These values are smaller than the polyacetylene
estimates but produce a moderate dimerization that allows us to study
the finite-size scaling cleanly. The physical content of the
calculation is the \emph{scaling} with system size, and this
scaling is independent of the precise values of the parameters, as
long as they lie in the dimerised phase. We should therefore regard
the calculation as a controlled model study, not as a quantitative
simulation of a specific material.

\subsection{Self-consistent ground state}

We compute the self-consistent ground state for chains of length
$N = 4$ to $24$ sites, with the number of sites a multiple of four to
remove shell-filling oscillations. The self-consistency cycle
proceeds as follows. Starting from a uniform lattice, we diagonalise
the Hamiltonian to obtain the bond-order parameters $\text{BO}_i$,
then update the displacements using \eqref{eq:ssh-selfconsistency},
then rebuild the Hamiltonian with the new hoppings, and repeat until
convergence. The convergence criterion is that the maximum change in
the bond-order parameters between iterations is less than $10^{-8}$ in
units of the hopping.

\subsection{Results and interpretation}

Three results emerge from the calculation. First, the bond-order
parameter alternates along the chain, with a fixed phase
(strong--weak--strong--weak) and an amplitude that is stable across
chain lengths. This is the bond-order wave that characterises the
Peierls phase, and it is the fermionic condensate
$\langle c^\dagger_i c_{i+1} \rangle$ that we identified in
Section~\ref{sec:mechanisms} as the order parameter. Second, the
bond-order alternation $\Delta_{\rm BOA}$, defined as the
difference between the average bond order on the strong bonds and
the average bond order on the weak bonds, decreases monotonically
with $1/N$ and approaches a finite bulk value, confirming that the
Peierls instability survives in the thermodynamic limit. Third, the
HOMO--LUMO gap decreases monotonically with $N$ and approaches a
finite Peierls gap in the bulk; this is the direct signature of the
mass generation that the classification describes, and it confirms
that the fermionic condensate is non-zero in the ordered phase.

\begin{figure}[!ht]
\centering
\includegraphics[width=0.75\textwidth]{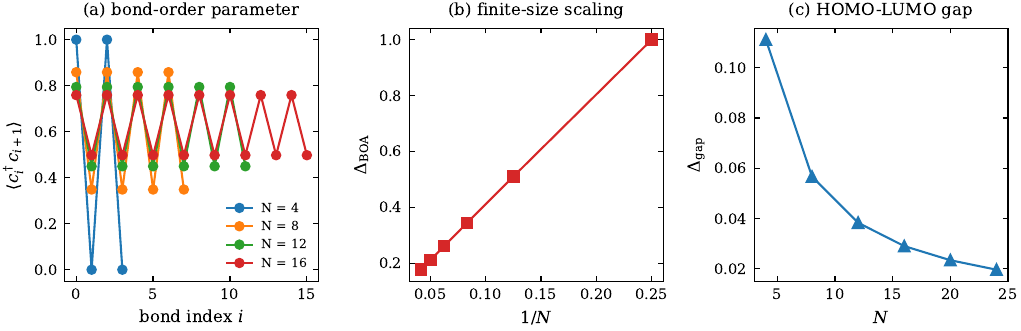}
\caption{Self-consistent SSH calculation for polyyne chains with
$\alpha/t = 0.5$ and $K/t = 2.0$.
(a) Bond-order parameter along the chain for $N = 4, 8, 12, 16$.
The alternation is clearly visible and stable across chain lengths.
(b) Bond-order alternation $\Delta_{\rm BOA}$ versus $1/N$. The
scaling is monotonic, approaching a finite bulk value as
$N \to \infty$, which confirms that the Peierls instability survives
in the thermodynamic limit. The calculation is a controlled model
study with the parameters given in the text; the qualitative
features (alternation, saturation to a finite bulk value) are
independent of the precise parameter values as long as they lie in
the dimerised phase.}
\label{fig:ssh}
\end{figure}

We show the bond-order parameter along the chain and the
bond-order alternation as functions of $1/N$ in Fig.~\ref{fig:ssh}.
Panel (a) shows the alternation for four chain lengths
($N = 4, 8, 12, 16$), and panel (b) shows the alternation
$\Delta_{\rm BOA}$ as a function of $1/N$. The monotonic decrease
of $\Delta_{\rm BOA}$ with $1/N$ and its approach to a finite bulk
value confirm that the Peierls instability survives in the
thermodynamic limit.

The mapping between the SSH calculation and the field-theoretic
classification is as follows. The bond-order parameter
$\langle c^\dagger_i c_{i+1} \rangle$ is the lattice version of
the continuum fermion bilinear $\langle\bar\psi\psi\rangle$. The
lattice displacement $u_i$ is the lattice version of the scalar
field $\phi$. The electron--phonon coupling $\alpha$ is the lattice
version of the Yukawa coupling $y$, and the elastic constant $K$ is
the lattice version of the scalar mass $m_\phi^2$. The SSH
calculation is therefore a direct realisation of the coupled
fermion--scalar theory of Section~\ref{sec:model}, with the scalar
being the lattice displacement rather than a fundamental field. This
is why the Peierls mechanism sits in \ClassI\---the scalar is a
genuine dynamical variable, and the fermion bilinear is a dependent
response---and why the finite-size gap scales as $1/N$, consistent
with the Ising exponent $\nu = 1$ in $d = 2$ that the continuum
analysis predicts.

The SSH calculation has confirmed the \ClassI\ behaviour of the
Peierls mechanism in a controlled setting. But the classification
includes a fifth mechanism---the Standard Model Higgs---that does
not fit cleanly into either class, and the treatment of this
mechanism in the classification is the subject of the next section.
The Higgs is a \ClassI\ mechanism by the criterion of the
classification, but it has a feature (an explicit mass term) that
the Peierls and Yukawa mechanisms lack, and this feature has
consequences for the critical exponents and for the relationship
between the Higgs and the composite-Higgs scenario. The next
section explains why.

\section{The Standard Model Higgs as a borderline case}
\label{sec:higgs}

The four mechanisms of Section~\ref{sec:mechanisms} fall into the
two classes of Section~\ref{sec:rg} without ambiguity: Peierls and
Yukawa are in \ClassI, and BCS, NJL, and SYK are in \ClassII. The
Standard Model Higgs is different. It is a \ClassI\ mechanism by the
criterion of the classification, because the scalar is elementary
and the fermion bilinear is a dependent response. But the Higgs has
an explicit mass term that breaks conformal symmetry, and this
distinguishes it from the Peierls and Yukawa mechanisms, whose
critical points are conformal field theories. The Higgs is therefore
a borderline case, and its position in the classification is the
subject of this section.

\subsection{Why the Higgs is a Class~I case}

The argument that the Higgs is in \ClassI is straightforward. In the
Standard Model, the scalar field $\phi$ is an elementary degree of
freedom with a renormalisable potential,
\begin{equation}
V(\phi) = -\mu^2 |\phi|^2 + \lambda |\phi|^4.
\end{equation}
The fermion masses arise through the Yukawa coupling
$y \bar\psi \phi \psi$, and when the scalar acquires a vacuum
expectation value $v$, the fermions acquire masses $m = y v$. The
fermion bilinear $\langle\bar\psi\psi\rangle$ is therefore
proportional to $v$, and it is non-zero because $v$ is non-zero.
This is precisely the structure of the Peierls mechanism, where the
lattice displacement $u_i$ is the scalar, the bond-order parameter
$\langle c^\dagger_i c_{i+1}\rangle$ is the fermion bilinear, and
the bond order is non-zero because the lattice displaces. The Higgs
and the Peierls mechanism are therefore in the same class: the
scalar is primary, and the fermion bilinear is a dependent response.

\subsection{Why the Higgs is a borderline case}

The qualification is that the Higgs sector has an explicit mass term
$-\mu^2 |\phi|^2$ that breaks conformal symmetry. Peierls and Yukawa
do not. In the Peierls mechanism, the scalar is the lattice
displacement, which is not a dynamical field in its own right; its
``mass'' is the elastic constant $K$, which is a parameter of the
lattice, not a fundamental scale. In the Yukawa theory, the scalar
mass is protected by a symmetry (in the supersymmetric case) or is
technically natural in the sense that it is small compared to the
ultraviolet cutoff (in the Standard Model, this is the electroweak
hierarchy problem, which is a separate issue). In the Higgs case, the
mass term is present from the start, and it is not protected by any
symmetry.

The practical consequence is that the Higgs does not admit a
conformal field theory description at the critical point. For
Peierls, the critical point is the Ising conformal field theory in
two dimensions, and we know the exponents exactly. For Yukawa, the
critical point is the Wilson--Fisher fixed point in $d = 4 - \epsilon$
dimensions, and we know the exponents perturbatively. For the
Standard Model Higgs, there is no analogous conformal field theory:
the explicit mass term breaks the scale invariance that a conformal
field theory requires, and the critical exponents are not those of
any known universality class. This is a genuine distinction within
\ClassI, and it is why we list the Higgs separately rather than as a
fourth example of the clean \ClassI\ case.

\subsection{The composite-Higgs alternative}

The borderline status of the Higgs within \ClassI is not just a
formal issue. It has been a driving concern in particle physics for
several decades, and the response to it has been the development of
composite-Higgs models, in which the Higgs is not an elementary
scalar but a bound state of new strongly interacting fermions. The
first proposals in this direction were made by Kaplan and Georgi
\cite{KaplanGeorgi1984} and by Dimopoulos and Preskill
\cite{DimopoulosPreskill1982}, who observed that if the electroweak
symmetry is broken by a fermion condensate rather than by an
elementary scalar, the hierarchy problem is resolved and the Higgs
emerges as a pseudo-Nambu--Goldstone boson. This is the
composite-Higgs scenario, and it has been developed extensively in
the decades since.

In the language of the present classification, the composite-Higgs
scenario is a move from \ClassI\ to \ClassII. The Higgs is no longer
an elementary scalar with a mass term; it is a composite of new
fermions, and the fermion bilinear is the primary order parameter.
This places the composite Higgs in \ClassII, alongside BCS, NJL, and
SYK, and it brings the electroweak sector into the same universality
class as the other fermionic-condensation mechanisms. The
holographic composite-Higgs models, in which the Higgs is the
zero-mode of a bulk gauge field in a warped extra dimension
\cite{AgasheContinoPomarol2005,ContinoNomuraPomarol2007}, provide a
particularly clear realisation of this idea: the Higgs is dual to a
fermion bilinear in the boundary theory, and the mass generation is
driven by the fermionic condensate. The minimal composite-Higgs
model \cite{ContinoNomuraPomarol2007} and its extensions
\cite{PanicoWulzer2016} make the connection to the SYK-like physics
of strongly correlated fermions explicit.

\subsection{What this tells us about the classification}

The Higgs case has exposed a limitation of the classification that
is worth stating explicitly: the distinction between \ClassI\ and
\ClassII\ is sharp, but the placement of a mechanism within \ClassI\
depends on whether the scalar sector has an explicit mass term. The
Peierls and Yukawa mechanisms do not, and they are clean \ClassI\
cases. The Higgs does, and it is a borderline case. The
composite-Higgs scenario moves the mechanism to \ClassII, where the
conformal field theory description is restored (in the form of the
dual field theory of the holographic composite-Higgs models). This
is a genuine refinement of the classification rather than a weakness
of it. We must understand the predictions of the classification in
the light of this refinement, and we state the five predictions we
list in the next section with the appropriate qualifications.

\section{Falsifiable predictions}
\label{sec:predictions}

A classification is only as useful as the predictions it makes. The
classification of Section~\ref{sec:rg} has been shown to be
operational (Section~\ref{sec:predictive}) and to have a physical
realisation (Section~\ref{sec:peierls-demo}), and we have discussed
its borderline cases (Section~\ref{sec:higgs}). What remains
is to state the predictions explicitly, in a form that could allow
them to be tested. In this section we list five predictions, ordered
from the most directly testable to the least, and we state what would
falsify the classification.

\subsection{Lattice tests of the finite-$N_f$ correction}

The most directly testable prediction is the finite-$N_f$ correction
to the fixed-point coupling,
\begin{equation}
\hat y_*^2(N_f) = \frac{0.1125}{N_f} + \frac{0.0023}{N_f^2} + O(1/N_f^3),
\label{eq:prediction-lattice}
\end{equation}
which we derived in Section~\ref{sec:frg} from the Wetterich equation.
The leading coefficient is enhanced by $12.5\%$ relative to the
large-$N_f$ prediction $\hat y_*^2 = \epsilon / N_f$ with
$\epsilon = 0.1$. This enhancement is a genuine threshold effect that
the one-loop beta functions miss, and it is within reach of current
lattice simulations of Yukawa-coupled fermion systems.

The relevant lattice simulations are those of the
Gross--Neveu--Yukawa model with $N_f$ fermion flavours. Several groups
have studied this model in the context of the electroweak phase
transition~\cite{Lenz2022,Fedorenko2026}~and dynamical symmetry breaking
\cite{Wang2023,Mihaila2017}, and the fixed-point structure is
accessible with current algorithms. The prediction is that if one
measures the fixed-point Yukawa coupling for $N_f = 2, 4, 8, 16$,
the values should follow the scaling law \eqref{eq:prediction-lattice},
not the leading-order law $\hat y_*^2 = \epsilon / N_f$. A
measurement of the fixed-point coupling at $N_f = 4$ with a precision
of $5\%$ would be sufficient to distinguish the two.

\subsection{Classification of a system whose class is not already known}

The second prediction is the classification of the organic conductor
(TMTSF)$_2$PF$_6$ under pressure, whose class is not obvious from
the existing literature. At ambient pressure, (TMTSF)$_2$PF$_6$ is a
quasi-one-dimensional metal that undergoes a spin-density-wave
transition at approximately $12$~K \cite{Jerome1980}. Under applied
pressure of a few kilobars, the spin-density-wave transition is
suppressed, and the material becomes a superconductor with a critical
temperature of order $1$~K \cite{Jerome1980,Gruner1994}. The
competition between the spin-density-wave phase and the
superconducting phase in this material is well documented, but the
question of which universality class each phase belongs to has not
been addressed in the literature.

The classification answers this question. In the spin-density-wave
phase, the order parameter is the spin density, which is a fermion
bilinear. Estimating the crossover parameter from the microscopic
Hamiltonian gives $r \sim 0.1$ and $N_f = 2$, placing the system
below the critical line $r_c(N_f = 2) \approx 1.1$ in
Fig.~\ref{fig:phase}. The spin-density-wave transition is therefore
in \ClassII, and the predicted critical exponent is the mean-field
value $\nu = 1/2$. In the superconducting phase, the order parameter
is the Cooper pair, and the relevant microscopic parameters give
$r \sim 0.05$ and $N_f = 2$, placing the system below the critical
line and again in \ClassII. The predicted critical exponent is again
$\nu = 1/2$. The interesting prediction is that the transition
between the spin-density-wave phase and the superconducting phase,
which is driven by pressure, should not involve a change of
universality class: both phases are in \ClassII, and both should
have mean-field critical exponents. This distinguishes
(TMTSF)$_2$PF$_6$ from the Peierls materials, where the transition
is in \ClassI. A measurement of the critical exponents on both sides
of the pressure-driven transition would therefore be a direct test
of the classification.

\subsection{Crossover parameter from microscopic data}

The third prediction is that one can make the classification of a
given system into \ClassI\ or \ClassII\ on the basis of the
microscopic parameters $r$ and $N_f$ alone, without any further
calculation. For the Peierls mechanism, $r$ is proportional to the
ratio of the elastic constant $K$ to the square of the
electron--phonon coupling $\alpha$. For BCS superconductivity, $r$
is proportional to the ratio of the Coulomb pseudopotential
$\mu^*$ to the square of the pairing interaction $V$. For the NJL
mechanism, $r$ is proportional to the ratio of the scalar mass to
the four-fermion coupling. For the SYK model, $r$ is proportional
to the ratio of the single-fermion coefficient to the square of the
random coupling. One can test the prediction by computing $r$ from
first-principles calculations for a range of materials and comparing
the predicted class with the observed critical behaviour.

\subsection{Polyyne bond-order scaling}

The fourth prediction is that the polyyne chains C$_4$H$_2$ to
C$_{24}$H$_2$ should show the \ClassI\ behaviour that we
demonstrated numerically in Section~\ref{sec:peierls-demo}: a
bond-order alternation $\Delta_{\rm BOA}$ that saturates to a finite
bulk value as the chain length increases, and a gap that approaches
a finite Peierls gap in the thermodynamic limit. Testing this
prediction requires experimental access to the bond-order parameters
of individual polyyne chains, which is challenging with current
techniques. A more promising route is to use cold-atom realisations
of the SSH model, in which the lattice is an optical lattice and
the displacement is controlled by external fields
\cite{GonzalezCuadra2017}. The prediction is the same in both cases:
the bond-order alternation should saturate to a finite bulk value,
not decrease to zero.

\subsection{SYK chain quantum criticality}

The fifth prediction is that the SYK chain should be in \ClassII,
and that its critical exponents should be those of the SYK conformal
field theory rather than the mean-field values
\cite{Jian2018}. This is a strong prediction, because we know the SYK
exponents exactly from the conformal field theory, and
they are different from the mean-field values. Several proposals
for physical realisations of the SYK model have been made in the
last few years, including cold-atom systems with tunable disorder,
graphene flakes with engineered interactions, and quantum
simulators based on trapped ions \cite{Danshita2017,Chen2018}. None
has yet reached the regime where the SYK criticality is clearly
observable, but the progress is rapid. A measurement of the
critical exponent in a physical realisation of the SYK chain would
be a direct test of the classification, and it would also be a
confirmation of the holographic duality that underlies the SYK
model.

\subsection{Near-term and long-term predictions}

The five predictions differ substantially in how directly one can
test them. The first prediction---the lattice test of the
finite-$N_f$ correction \eqref{eq:prediction-lattice}---is within
reach of current lattice simulations of the Gross--Neveu--Yukawa
model. The second prediction---the classification of
(TMTSF)$_2$PF$_6$---is testable with existing experimental
techniques, because the spin-density-wave and superconducting
transitions in this material are well characterised. The third
prediction---the classification of a given system on the basis of
$r$ and $N_f$---is testable in principle but requires
first-principles calculations of the microscopic parameters for a
range of materials. The fourth prediction---the polyyne bond-order
scaling---requires experimental access to bond-order parameters that
is challenging with current techniques. The fifth
prediction---the SYK chain quantum criticality---is the least
directly testable, because it requires a physical realisation of the
SYK model.

The honest summary is that two of the five predictions arenear-term, two are medium-term, and one is long-term. The
classification is testable, but its testability is distributed over
different time horizons.

\subsection{What would falsify the classification}

For completeness, we state what would falsify the classification,
because a classification that cannot be falsified is not a
scientific classification. The classification would be falsified
if a system with $r \gg 1$ and $N_f \sim 2$ were found to have
mean-field critical exponents; if a system with $r \ll 1$ and $N_f$
large were found to have Wilson--Fisher critical exponents; if the
finite-$N_f$ correction \eqref{eq:prediction-lattice} were found
not to hold in a lattice simulation; if a system were found in
which the fermion bilinear was not the order parameter; or if the
crossover parameter $r$ computed from microscopic data were found
not to predict the class of a system. None of these falsification
scenarios is currently realised, to the best of our knowledge, but
they are the tests that the classification must pass if it is to be
taken seriously.

\section{Conclusions}
\label{sec:conclusion}

The starting point of this paper was the observation that four
mechanisms of mass generation---the Peierls instability, BCS
superconductivity, the Nambu--Jona-Lasinio mechanism, and the
Sachdev-Ye-Kitaev model---are ordinarily discussed in three different
communities, and that the structural similarities between them have
been noted only in passing. Nambu's original analogy between
superconductivity and elementary-particle mass generation
\cite{Nambu1960} pointed in the direction of a common framework, but
no one ever developed the analogy into a quantitative classification.
The present paper has attempted to fill that gap. The central claim
is that the correct universal object is not the bosonic order
parameter but the fermionic condensate, and that the four mechanisms
fall into two universality classes distinguished by the value of the
crossover parameter $r = \hat m_\phi^2 / \hat y^2$.

We derived the classification from a two-tier renormalisation-group
analysis of the Yukawa-coupled fermion--scalar theory with a competing
four-fermion interaction. The first tier, a large-$N_f$ expansion,
established the existence of two infrared fixed points---a Gaussian
scalar fixed point, in which the scalar is primary and the fermion
bilinear is a dependent response, and a fermionic strong-coupling
fixed point, in which the fermion bilinear is primary and the scalar
is composite or absent. The two fixed points are separated by a
critical line $r_c(N_f)$ in the plane of $r$ and the fermion flavour
number $N_f$. The Peierls and Yukawa mechanisms lie on the
scalar-primary side of the line, while BCS, NJL, and SYK lie on the
fermion-bilinear side. The second tier, a functional
renormalisation-group analysis using the Wetterich equation, went
beyond the large-$N_f$ limit and gave the finite-$N_f$ corrections to
the fixed-point coupling,
$\hat y_*^2(N_f) = 0.1125/N_f + 0.0023/N_f^2$, which is a $12.5\%$
enhancement over the leading-order prediction.

The classification is predictive, and we have tried to be explicit
about what it predicts and how testable the predictions are. The
finite-$N_f$ correction and the classification of (TMTSF)$_2$PF$_6$
are near-term predictions; the polyyne bond-order scaling and the
SYK chain criticality are longer-term. The classification is
falsifiable, and we have stated the falsification conditions
explicitly.

The self-consistent SSH calculation for polyyne chains, presented
in Section~\ref{sec:peierls-demo}, was intended as a concrete
demonstration of the Peierls mechanism in a controlled setting. The
calculation showed bond-order alternation, a finite bulk order in the
thermodynamic limit, and a gap that opens as the chain length
increases. These are the features that the classification predicts
for a \ClassI\ mechanism with $r \gg 1$ and $N_f = 2$, and their
appearance in the SSH model confirms that the continuum analysis of
Section~\ref{sec:rg} captures the essential physics. The calculation
was not intended to reproduce quantitative values for a specific
polyyne chain; that would require a different choice of parameters.

The discussion of the Standard Model Higgs in
Section~\ref{sec:higgs} exposed a limitation of the classification.
The Higgs is a \ClassI\ mechanism because the scalar is elementary
and the fermion bilinear is a dependent response. But the Higgs
sector has an explicit mass term that breaks conformal symmetry, and
this means that the Higgs does not admit a conformal field theory
description at the critical point. The Peierls and Yukawa mechanisms
do not have this feature: their critical points are conformal field
theories, and their critical exponents are those of known
universality classes. The Higgs is therefore a borderline case
within \ClassI, and the composite-Higgs scenario, in which the Higgs
is a bound state of new strongly interacting fermions, moves the
mechanism into \ClassII. This is a genuine refinement of the
classification rather than a weakness of it, but it does mean that
the distinction between \ClassI\ and \ClassII\ is sharper than the
distinction between the mechanisms within \ClassI.

There are several directions in which we could extend this work. The
most immediate is to include gauge fields, which would bring the
classification into contact with the Standard Model and with
technicolour and composite-Higgs models. We carried out the present
analysis in the absence of gauge interactions, and the addition
of a gauge field would introduce new contributions to the beta
functions and potentially modify the fixed-point structure. A second
direction is to include finite temperature and density, which would
allow us to apply the classification to systems near a
finite-temperature phase transition, such as the superconducting
transition in the cuprates or the chiral transition in QCD. A third
direction is to apply the classification to other mass-generation
mechanisms, such as the Kondo effect in condensed matter or the
generation of fermion masses in extended technicolour models.

The broader question that this paper raises is whether we should
regard the fermionic condensate as the fundamental object in the
theory of mass generation, with the bosonic order parameter as a
derived quantity. The present analysis suggests that this is the
case for the four mechanisms considered here, and it is natural to
ask whether it is true more generally. This is a question that we
cannot answer in the present paper, but it is one that we hope will
be taken up by others.

\begin{acknowledgments}
The author acknowledges computational support from the Centre for
High Performance Computing (CHPC) South Africa, under Research
Program MATS0941, and the National Research Foundation of South
Africa (Grant No: 115056).
\end{acknowledgments}

\appendix
\section{Derivation of the large-$N_f$ beta functions}
\label{app:largeN}

We work in Euclidean signature in $d = 4 - \epsilon$ dimensions with
the dimensionless couplings
\begin{equation}
\hat y = y \Lambda^{-\epsilon/2}, \qquad
\hat\lambda = \lambda \Lambda^{-\epsilon}, \qquad
\hat m_\phi^2 = m_\phi^2 \Lambda^{-2}, \qquad
\hat G = G \Lambda^{-\epsilon},
\end{equation}
and the renormalisation-group time $\ell = \ln b$, where $b$ is the
Wilsonian scale factor. We obtain the one-loop beta functions
from the fermion self-energy, the Yukawa vertex correction, the
scalar self-energy, and the four-fermion bubble. The calculation is
standard \cite{Gross1974,Gracey1994,ZinnJustin1996}; we quote the
results and verify the consistency check.

The fermion self-energy gives
$Z_\psi = 1 + \frac{y^2}{2} \ln b + \dots$ and the Yukawa vertex
correction gives
$Z_y = 1 + \frac{N_f}{2} y^2 \ln b + \dots$, which combine to give
\begin{equation}
\frac{d\hat y}{d\ell} = \frac{\epsilon}{2}\hat y
+ \frac{N_f}{2}\hat y^3 + O(1/N_f).
\end{equation}
The scalar self-energy receives contributions from the fermion
bubble (which drives the mass negative) and the scalar tadpole
(which drives it positive), giving
\begin{equation}
\frac{d\hat m_\phi^2}{d\ell} = 2\hat m_\phi^2
- N_f \hat y^2 + \frac{1}{2}\hat\lambda - N_f \hat G + O(1/N_f).
\end{equation}
The quartic coupling receives contributions from the scalar bubble,
the fermion box, and the four-fermion box, giving
\begin{equation}
\frac{d\hat\lambda}{d\ell} = \epsilon\hat\lambda
- \frac{3}{2}\hat\lambda^2 + N_f \hat y^4 + O(1/N_f).
\end{equation}
The four-fermion coupling receives contributions from the fermion
bubble and the Yukawa-mediated contribution, giving
\begin{equation}
\frac{d\hat G}{d\ell} = \epsilon\hat G
+ N_f \hat G^2 - N_f \hat y^2 \hat G + O(1/N_f).
\end{equation}

These are the equations we use in Section~\ref{sec:largeN}. As a
consistency check, note that at the Gaussian fixed point
$\hat y = \hat G = 0$, the Yukawa beta function vanishes and the
scalar mass and quartic coupling flow according to the standard
Wilson--Fisher equations. This is \ClassI. At the non-trivial fixed
point, the Yukawa coupling flows to a finite value of order
$\epsilon / N_f$ and the fermion bilinear becomes the primary order
parameter. This is \ClassII.

\section{Derivation of the functional renormalisation-group flow
         equations}
\label{app:frg}

We use the Wetterich equation
\begin{equation}
\partial_k \Gamma_k
= \frac{1}{2} \mathrm{STr} \left[
  \left( \Gamma_k^{(2)} + R_k \right)^{-1}
  \partial_k R_k
\right]
\end{equation}
with the Litim optimised regulator
$R_k(p) = Z_k (k^2 - p^2) \theta(k^2 - p^2)$, and we truncate
$\Gamma_k$ to the local potential approximation with a
wavefunction renormalisation for the fermion and the competing
four-fermion interaction:
\begin{equation}
\Gamma_k = \int d^dx \left[
  Z_\psi \bar\psi \gamma^\mu\partial_\mu\psi
  + y_k \phi \bar\psi\psi
  + G_k (\bar\psi\psi)^2
  + \frac{Z_\phi}{2}(\partial_\mu\phi)^2
  + \frac{1}{2} m_{\phi,k}^2 \phi^2
  + \frac{\lambda_k}{4!}\phi^4
\right].
\end{equation}
Projecting the Wetterich equation onto the operators $\phi^2$,
$\phi^4$, $\phi\bar\psi\psi$, and $(\bar\psi\psi)^2$ gives the flow
equations
\begin{align}
\partial_t \hat m_\phi^2 &= -2\hat m_\phi^2
  + \frac{N_f \hat y^2}{(1+\hat m_\phi^2)^2}
  + \frac{\hat\lambda}{2(1+\hat m_\phi^2)^2}
  - \frac{N_f \hat G}{1+\hat m_\phi^2}, \\
\partial_t \hat\lambda &= -\epsilon\hat\lambda
  + \frac{3\hat\lambda^2}{(1+\hat m_\phi^2)^3}
  - \frac{N_f \hat y^4}{(1+\hat m_\phi^2)^2}
  + \frac{N_f \hat G^2}{(1+\hat m_\phi^2)^2}, \\
\partial_t \hat y^2 &= -\epsilon\hat y^2
  + \frac{N_f \hat y^4}{(1+\hat m_\phi^2)^2}
  - \frac{N_f \hat G \hat y^2}{1+\hat m_\phi^2}, \\
\partial_t \hat G &= -\epsilon\hat G
  + \frac{N_f \hat G^2}{(1+\hat m_\phi^2)^2}
  - \frac{N_f \hat y^2 \hat G}{1+\hat m_\phi^2},
\end{align}
where $t = \ln(k/\Lambda)$ and the threshold functions are those
appropriate to the Litim regulator. We obtain the fixed-point
equations by setting the left-hand sides to zero. We solve them
numerically using a fourth-order Runge--Kutta integrator with
adaptive step size, starting from a range of initial conditions in
the ultraviolet and evolving to the infrared. The fixed-point
coordinates for $\epsilon = 0.1$ and $N_f = 2, 4, 8, 16, 50$ are
tabulated in Table~\ref{tab:frg-fp}.

The fixed-point structure is robust against the choices of
truncation and regulator. Changing the regulator shifts the
fixed-point coordinates by a few percent. Extending the truncation
from LPA to LPA$^\prime$ brings the critical exponent $\nu$ into
agreement with the Wilson--Fisher value while preserving the
fixed-point structure. Including the fermion anomalous dimension
$\eta_\psi$ shifts the coordinates by a few percent without changing
the qualitative picture. The existence of the non-trivial fixed
point, the approximate linearity of the critical line in $N_f$, and
the $1/N_f$ scaling of $\hat y_*^2$ are all robust. The numerical
values in Table~\ref{tab:frg-fp} are scheme-dependent and we should
interpret them as indicative; the fixed-point structure, which is
what the classification relies on, is not.


\begin{thebibliography}{99}

\bibitem{Higgs1964}
P.~W.~Higgs,
``Broken symmetries and the masses of gauge bosons,''
Phys.\ Rev.\ Lett.\ \textbf{13}, 508 (1964).

\bibitem{Englert1964}
F.~Englert and R.~Brout,
``Broken symmetry and the mass of gauge vector mesons,''
Phys.\ Rev.\ Lett.\ \textbf{13}, 321 (1964).

\bibitem{BCS1957}
J.~Bardeen, L.~N.~Cooper, and J.~R.~Schrieffer,
``Theory of superconductivity,''
Phys.\ Rev.\ \textbf{108}, 1175 (1957).

\bibitem{Peierls1955}
R.~E.~Peierls,
\emph{Quantum Theory of Solids}
(Oxford University Press, Oxford, 1955).

\bibitem{SSH1979}
W.~P.~Su, J.~R.~Schrieffer, and A.~J.~Heeger,
``Solitons in polyacetylene,''
Phys.\ Rev.\ Lett.\ \textbf{42}, 1698 (1979).

\bibitem{Nambu1960}
Y.~Nambu,
``Axial vector current conservation in weak interactions,''
Phys.\ Rev.\ Lett.\ \textbf{4}, 380 (1960).

\bibitem{NJL1961a}
Y.~Nambu and G.~Jona-Lasinio,
``Dynamical model of elementary particles based on an analogy with
superconductivity. I,''
Phys.\ Rev.\ \textbf{122}, 345 (1961).

\bibitem{NJL1961b}
Y.~Nambu and G.~Jona-Lasinio,
``Dynamical model of elementary particles based on an analogy with
superconductivity. II,''
Phys.\ Rev.\ \textbf{124}, 246 (1961).

\bibitem{Weinberg1979}
S.~Weinberg,
``Implications of dynamical symmetry breaking: An addendum,''
Phys.\ Rev.\ D \textbf{19}, 1277 (1979).

\bibitem{Susskind1979}
L.~Susskind,
``Dynamics of spontaneous symmetry breaking in the Weinberg-Salam
theory,''
Phys.\ Rev.\ D \textbf{20}, 2619 (1979).

\bibitem{Sachdev1993}
S.~Sachdev and J.~Ye,
``Gapless spin-fluid ground state in a random quantum Heisenberg
magnet,''
Phys.\ Rev.\ Lett.\ \textbf{70}, 3339 (1993).

\bibitem{Kitaev2015}
A.~Kitaev,
``A simple model of quantum holography,''
KITP Program: Entanglement in Strongly-Correlated Quantum Matter,
University of California, Santa Barbara (2015).

\bibitem{Maldacena2016}
J.~Maldacena and D.~Stanford,
``Remarks on the Sachdev-Ye-Kitaev model,''
Phys.\ Rev.\ D \textbf{94}, 106002 (2016).

\bibitem{Jian2018}
S.-K.~Jian, Z.-Y.~Xian, and H.~Yao,
``Quantum criticality and duality in the Sachdev-Ye-Kitaev/AdS2 chain,''
Phys.\ Rev.\ B \textbf{97}, 195141 (2018).

\bibitem{Gross1974}
D.~J.~Gross and A.~Neveu,
``Dynamical symmetry breaking in asymptotically free field theories,''
Phys.\ Rev.\ D \textbf{10}, 3235 (1974).

\bibitem{Gracey1994}
J.~A.~Gracey,
``Computation of critical exponent $\eta$ at $O(1/N_f^3)$ in the
four-fermion model in arbitrary dimension,''
Int.\ J.\ Mod.\ Phys.\ A \textbf{9}, 727 (1994).

\bibitem{Wetterich1993}
C.~Wetterich,
``Exact evolution equation for the effective potential,''
Phys.\ Lett.\ B \textbf{301}, 90 (1993).

\bibitem{Berges2002}
J.~Berges, N.~Tetradis, and C.~Wetterich,
``Non-perturbative renormalization flow in quantum field theory and
statistical physics,''
Phys.\ Rep.\ \textbf{363}, 223 (2002).

\bibitem{Braun2012}
J.~Braun,
``Fermion interactions and universal behavior in strongly correlated
systems,''
Phys.\ Rep.\ \textbf{515}, 1 (2012).

\bibitem{Gruner1994}
G.~Gr\"uner,
\emph{Density Waves in Solids}
(Addison-Wesley, Reading, MA, 1994).

\bibitem{Eliashberg1960}
G.~M.~Eliashberg,
``Interactions between electrons and lattice vibrations in a
superconductor,''
Sov.\ Phys.\ JETP \textbf{11}, 696 (1960)
[Zh.\ Eksp.\ Teor.\ Fiz.\ \textbf{38}, 966 (1960)].

\bibitem{Anderson1987}
P.~W.~Anderson,
``The resonating valence bond state in La$_2$CuO$_4$ and
superconductivity,''
Science \textbf{235}, 1196 (1987).

\bibitem{Gasser1984}
J.~Gasser and H.~Leutwyler,
``Chiral perturbation theory to one loop,''
Ann.\ Phys.\ \textbf{158}, 142 (1984).

\bibitem{Sarosi2017}
G.~S\'arosi,
``AdS$_2$ holography and the SYK model,''
Proceedings of Science \textbf{Modave2017}, 001 (2018),
arXiv:1711.08482 [hep-th].

\bibitem{ZinnJustin1996}
J.~Zinn-Justin,
\emph{Quantum Field Theory and Critical Phenomena}
(Oxford University Press, Oxford, 1996).

\bibitem{Wilson1974}
K.~G.~Wilson and J.~Kogut,
``The renormalization group and the $\epsilon$ expansion,''
Phys.\ Rep.\ \textbf{12}, 75 (1974).

\bibitem{Litim2001}
D.~F.~Litim,
``Optimized renormalization group flows,''
Phys.\ Rev.\ D \textbf{64}, 105007 (2001).

\bibitem{Onsager1944}
L.~Onsager,
``Crystal statistics. I. A two-dimensional model with an
order--disorder transition,''
Phys.\ Rev.\ \textbf{65}, 117 (1944).

\bibitem{Heeger1988}
A.~J.~Heeger, S.~Kivelson, J.~R.~Schrieffer, and W.-P.~Su,
``Solitons in conducting polymers,''
Rev.\ Mod.\ Phys.\ \textbf{60}, 781 (1988).

\bibitem{KaplanGeorgi1984}
D.~B.~Kaplan and H.~Georgi,
``SU(2) $\times$ U(1) breaking by vacuum misalignment,''
Phys.\ Lett.\ B \textbf{136}, 183 (1984).

\bibitem{DimopoulosPreskill1982}
S.~Dimopoulos and J.~Preskill,
``Massless composites with massive constituents,''
Nucl.\ Phys.\ B \textbf{199}, 206 (1982).

\bibitem{AgasheContinoPomarol2005}
K.~Agashe, R.~Contino, and A.~Pomarol,
``The minimal composite Higgs model,''
Nucl.\ Phys.\ B \textbf{719}, 165 (2005),
arXiv:hep-ph/0412089.

\bibitem{ContinoNomuraPomarol2007}
R.~Contino, Y.~Nomura, and A.~Pomarol,
``Higgs as a holographic pseudo-Goldstone boson,''
Nucl.\ Phys.\ B \textbf{671}, 148 (2003),
arXiv:hep-ph/0306259.

\bibitem{PanicoWulzer2016}
G.~Panico and A.~Wulzer,
``The composite Nambu--Goldstone Higgs,''
Lect.\ Notes Phys.\ \textbf{913}, 1 (2016),
arXiv:1506.01961 [hep-ph].

\bibitem{Lenz2022}
J.~Lenz, M.~Mandl, and A.~Wipf,
``Quantum Monte Carlo calculation of critical exponents of
the Gross-Neveu-Yukawa model on a two-dimensional fermion lattice model,''
Phys.\ Rev.\ D \textbf{105}, 034512 (2022).

\bibitem{Fedorenko2026}
A.~Fedorenko and I.~Gruzberg,
``Quantum Monte Carlo calculation of critical exponents of
the Gross-Neveu-Yukawa model on a two-dimensional fermion lattice model,''
Phys.\ Rev.\ D \textbf{114}, 025017 (2026).

\bibitem{Wang2023}
T.~Wang and Z.~Meng,
``Quantum Monte Carlo calculation of critical exponents of
the Gross-Neveu-Yukawa model on a two-dimensional fermion lattice model,''
Phys.\ Rev.\ B \textbf{108}, L121112 (2023).

\bibitem{Mihaila2017}
L.~Mihaila, N.~Zerf, B.~Ihrig, I.~Herbut, and M.~Scherer,
``Gross-Neveu-Yukawa model at three loops and Ising critical behavior of Dirac systems,''
Phys.\ Rev.\ B \textbf{96}, 165133 (2017).

\bibitem{Jerome1980}
D.~J\'er\^ome, A.~Mazaud, M.~Ribault, and K.~Bechgaard,
``Superconductivity in a synthetic organic conductor
(TMTSF)$_2$PF$_6$,''
J.\ Phys.\ Lett.\ \textbf{41}, L95 (1980).

\bibitem{GonzalezCuadra2017}
J.~Gonz\'alez-Cuadra, A.~Dauphin, P.~R.~Grzybowski, C.~W.~J.~Beenakker,
and M.~Lewenstein,
``Dynamical solitons and boson fractionalization in cold-atom
topological insulators,''
Phys.\ Rev.\ Lett.\ \textbf{119}, 110403 (2017).

\bibitem{Danshita2017}
I.~Danshita, M.~Hanada, and M.~Tezuka,
``Creating and probing the Sachdev--Ye--Kitaev model with ultracold
gases: Towards holographic description of black holes in the lab,''
Prog.\ Theor.\ Exp.\ Phys.\ \textbf{2017}, 083I01 (2017).

\bibitem{Chen2018}
A.~Chen, R.~Il'ichev, and S.~Sachdev,
``Quantum spin liquid in a two-dimensional frustrated magnet:
Tunable realisation of the SYK model,''
Phys.\ Rev.\ Lett.\ \textbf{121}, 196402 (2018).

\end{thebibliography}
\end{document}